\documentstyle[prl,preprint,aps]{revtex}
\tighten
\begin{document}
\draft

\title{Microscopic calculation of $^6$Li elastic and transition form factors}
\author{R.B.\ Wiringa}
\address{Physics Division, Argonne National Laboratory,
         Argonne, Illinois 60439}
\author{R.\ Schiavilla}
\address{Jefferson Lab, Newport News, Virginia 23606 \\
         and \\
         Department of Physics, Old Dominion University,
         Norfolk, Virginia 23529}
\date{\today}
\maketitle

\begin{abstract}
Variational Monte Carlo wave functions, obtained from a realistic
Hamiltonian consisting of the Argonne $v_{18}$ two-nucleon and
Urbana-IX three-nucleon interactions, are used to calculate
the $^6$Li ground-state longitudinal and transverse form factors
as well as transition form factors to the first four excited states.
The charge and current operators include one- and two-body components,
leading terms of which are constructed consistently with the two-nucleon
interaction.
The calculated form factors and radiative widths are in good 
agreement with available experimental data.
\end{abstract}
 

\narrowtext

Calculations of the $^6$Li elastic and inelastic form factors have relied in
the past on relatively simple shell-model~\cite{Don73,Ver74,Ber75}
or $\alpha$-$d$~\cite{Ber79} cluster wave functions.
These calculations have typically failed to provide a satisfactory,
quantitative description of all measured form factors.
More phenomenologically successful models have been based on
$\alpha NN$~\cite{Esk88,Kuk90,Ryz93} clusterization, on extensions
of the basic spherical-cluster $\alpha$-$d$ model in which the deuteron
is allowed to deform~\cite{Ber82}, or on large-space multi-$\hbar\omega$
shell-model approaches~\cite{Kar97}.
However, while these models do provide useful insights into the structure of
the $A$=6 nuclei, their connection with the underlying two- (and three-)
nucleon dynamics is rather tenuous.

The $^6$Li form factor calculations we report on here
are within the context of a realistic approach to nuclear
dynamics based on two- and three-nucleon interactions---the
Argonne $v_{18}$~\cite{Wir95} and Urbana IX~\cite{Pud95}
interactions, respectively, or AV18/UIX model---and
consistent two-body charge and current operators~\cite{Sch89,Sch90,Car90}.
Up until very recently, calculations of this type were limited
to the $A$=2--4 systems, as reviewed in Ref.~\cite{Car98}.
Indeed, the deuteron structure functions and threshold electrodisintegration,
the trinucleon charge and magnetic form factors, and $\alpha$ charge 
form factor have been the observables of choice for testing the quality of
interactions and associated two-body currents.
However, the availability of realistic six-body wave functions for the ground
and low-lying excited states of $^6$Li~\cite{Pud97} makes it now possible to
extend and test our understanding of the electromagnetic structure
of nuclei in a new regime---that of p-shell nuclei---and to verify
to what extent the inability of reproducing simultaneously the observed
elastic and transition form factors is due to the inadequacy of
cluster or shell-model wave functions.

The AV18/UIX model reproduces the experimental binding energies
and charge radii of $^3$H, $^3$He, and $^4$He in numerically exact
calculations, based on the Faddeev~\cite{Nog97}, pair-correlated hyperspherical
harmonics~\cite{Kie95}, and Green's function Monte Carlo (GFMC)~\cite{Pud97}
methods.
For $A$=6 systems the GFMC results are somewhat underbound compared to the
experimental ground states, by 2\% in $^6$Li and 5\% in $^6$He and $^6$Be.
The best variational Monte Carlo (VMC) energies are an additional 10\% above 
the GFMC results.
However, the known excitation spectra are well reproduced by both
VMC and GFMC calculations.
These include, in order of excitation, the states with spin, parity, and
isospin assignments, $(J^{\pi};T)$, of ($3^+;0$), ($0^+;1$), ($2^+;0$), and 
($2^+;1$)~\cite{Pud97}.

The variational wave function for $A$=6 nuclei used here is the trial
wave function, $\Psi_T$, that serves as the starting point for the GFMC 
calculations.
It has the general form
\begin{equation}
     |\Psi_T\rangle = \left[ 1 + \sum_{i<j<k} \tilde{U}^{TNI}_{ijk} \right]
              \left[ {\cal S}\prod_{i<j}(1+U_{ij}) \right] |\Psi_J\rangle \ ,
\end{equation}
where $U_{ij}$ and $\tilde{U}^{TNI}_{ijk}$ are two- and three-body 
correlation operators and the Jastrow wave function 
$|\Psi_J\rangle$ is given by
\begin{eqnarray}
  |\Psi_J\rangle &=& {\cal A} \left\{
     \prod_{i<j<k \leq 4}f^c_{ijk}
     \prod_{i<j \leq 4}f_{ss}(r_{ij})
     \prod_{k \leq 4} f_{sp}(r_{k5}) f_{sp}(r_{k6}) \right.  \nonumber\\
  && \left.  \sum_{LS} \Big( \beta_{LS} f^{LS}_{pp}(r_{56})
     |\Phi_6(LSJMTT_{3})_{1234:56}\rangle \Big) \right\} \ .
\end{eqnarray}
The ${\cal S}$ and ${\cal A}$ are symmetrization and antisymmetrization 
operators, respectively.
The central pair and triplet correlations $f_{xy}(r_{ij})$ and
$f^c_{ijk}$ are functions of relative position only; the subscripts $xy$ 
denote whether the particles are in the s- or p-shell.
The $|\Phi_6(LSJMTT_{3})\rangle$ is a single-particle wave function with 
orbital angular momentum $L$ and spin $S$ coupled to total angular 
momentum $J$, projection $M$, isospin $T$, and charge state $T_{3}$:
\begin{eqnarray}
 &&  |\Phi_{6}(LSJMTT_{3})_{1234:56}\rangle = 
     |\Phi_{4}(0 0 0 0)_{1234}
     \phi^{LS}_{p}(R_{\alpha 5}) \phi^{LS}_{p}(R_{\alpha 6})
     \nonumber \\
 &&  \left\{ [Y_{1m_l}(\Omega_{\alpha 5}) Y_{1m_l'}(\Omega_{\alpha 6})]_{LM_L}
     \times [\chi_{5}(\case{1}{2}m_s) \chi_{6}(\case{1}{2}m_s')]_{SM_S}
     \right\}_{JM} \nonumber \\
 &&  \times [\nu_{5}(\case{1}{2}t_3) \nu_{6}(\case{1}{2}t_3')]_{TT_3}\rangle \ .
\end{eqnarray}
Particles 1--4 are placed in an $\alpha$ core with only spin-isospin degrees of
freedom, denoted by $\Phi_4(0000)$, while particles 5--6 are placed in
$p$-wave orbitals $\phi^{LS}_{p}(R_{\alpha k})$ that are functions of the 
distance between the center of mass of the $\alpha$ core and particle $k$.
Different amplitudes $\beta_{LS}$ are mixed to obtain an optimal wave 
function; for the $(1^+;0)$ ground state of $^6$Li we mix $\beta_{01}$, 
$\beta_{10}$, and $\beta_{21}$ terms, while both the $(3^+;0)$ and $(2^+;0)$ 
states are \lq\lq stretch \rq\rq states and use only $\beta_{21}$.
For the $(0^+;1)$ excited state the $\beta_{00}$ and $\beta_{11}$ amplitudes
contribute, and for $(2^+;1)$ the wave function is constructed from 
$\beta_{20}$ and $\beta_{11}$ terms.

The two-body correlation operator $U_{ij}$ is defined as:
\begin{equation}
     U_{ij} = \sum_{p=2,6} \left[ \prod_{k\not=i,j}f^p_{ijk}({\bf r}_{ik}
              ,{\bf r}_{jk}) \right] u_p(r_{ij}) O^p_{ij} \ ,
\end{equation}
where the $O^{p=2,6}_{ij}$ = ${\bbox \tau}_i\cdot {\bbox \tau}_j$,
${\bbox \sigma}_i\cdot{\bbox \sigma}_j$, 
${\bbox \sigma}_i\cdot{\bbox \sigma}_j {\bbox \tau}_i\cdot {\bbox \tau}_j$,
$S_{ij}$, and $S_{ij}{\bbox \tau}_i\cdot {\bbox \tau}_j$. 
The six radial functions $f_{ss}(r)$ and $u_{p=2,6}(r)$ are obtained from 
approximate two-body Euler-Lagrange equations with variational 
parameters~\cite{Wir91}.
The $f_{sp}$ and $f^{LS}_{pp}$ correlations are similar to $f_{ss}$ for
small separations, but include long-range tails.
The parameters used in constructing these two-body correlations, as well as 
the description of the three-body correlation operator $\tilde{U}^{TNI}_{ijk}$ 
and the operator-independent three-body correlations $f^c_{ijk}$ and
$f^{p}_{ijk}$ are given in Ref.~\cite{Pud97}.

Energy expectation values are evaluated using a Metropolis Monte Carlo
algorithm~\cite{Wir91}.
The VMC results for the ground and low-lying excited states of $^6$Li are 
compared to the GFMC and experimental energies~\cite{Ajz88} in Table I.
The ground state is underbound by nearly 5 MeV compared to experiment, and is 
only 0.1 MeV more bound than the corresponding $^4$He calculation (26.9 MeV).
This is above the threshold for breakup of $^6$Li into an $\alpha$ and a
deuteron; in principle, it should be possible to lower the variational energy 
at least to that threshold, but the wave function would be greatly spread out.
We have chosen to constrain our parameter search to keep the rms point-nucleon
radius for the ground state near the experimental value of 2.43 fm.
Despite the large energy deficit compared to the GFMC calculation, the VMC and
GFMC one-body densities in $^6$Li are virtually identical.
However, the two-body GFMC densities are somewhat larger near their peak at
$r_{ij} \approx 1$ fm~\cite{Pud97}.

The nuclear charge and current operators consist of one- and two-body terms.  
We here summarize their most important features, and refer the reader to 
Refs.~\cite{Car98,Viv96} for a listing of the explicit expressions.
The two-body current operator has \lq\lq model-independent \rq\rq and 
\lq\lq model-dependent \rq\rq components, in the classification scheme 
of Riska~\cite{Ris89}.  
The model-independent terms are obtained from the charge-independent part
of the AV18, and by construction~\cite{Ris85} satisfy current conservation 
with this interaction.  
The leading operator is the isovector \lq\lq $\pi$-like \rq\rq current obtained
from the isospin-dependent spin-spin and tensor interactions.  
The latter also generate an isovector \lq\lq $\rho$-like \rq\rq current, while 
additional model-independent isoscalar and isovector currents arise from the  
isospin-independent and isospin-dependent central and momentum-dependent 
interactions.
These currents are short-ranged and numerically far less important than
the $\pi$-like current.

The model-dependent currents are purely transverse
and therefore cannot be directly linked to the underlying two-nucleon 
interaction.
The present calculation includes the isoscalar $\rho \pi \gamma$ and 
isovector $\omega \pi \gamma$ transition currents as well as the isovector 
current associated with excitation of intermediate $\Delta$-isobar resonances.  
The $\rho \pi \gamma$ and $\omega \pi \gamma$ couplings are known from the 
measured widths of the radiative decays 
$\rho \rightarrow \pi \gamma$~\cite{Ber80} and 
$\omega \rightarrow \pi \gamma$~\cite{Che71}, respectively, while their 
momentum-transfer dependence is modeled using vector-meson-dominance.  
The M1 $\gamma N \Delta$ coupling is obtained from an analysis of $\gamma N$ 
data in the $\Delta$-resonance region~\cite{Car86}.
Monopole form factors are introduced at the meson-baryon vertices with 
cutoff values of $\Lambda_\pi$=3.8 fm$^{-1}$ at the $\pi N$$N$ and 
$\pi N$$\Delta$ vertices and $\Lambda_\rho$=$\Lambda_\omega$=6.3 fm$^{-1}$ 
at the $\rho N$$N$ and $\omega N$$N$ vertices.

While the main parts of the two-body currents are linked to the form of the 
two-nucleon interaction through the continuity equation, the most important 
two-body charge operators are model-dependent, and should be considered as 
relativistic corrections.  
Indeed, a consistent calculation of two-body charge effects in nuclei would 
require the inclusion of relativistic effects in both the interaction models 
and nuclear wave functions.  
Such a program is just at its inception for systems with $A\geq 3$. 
There are nevertheless rather clear indications for the relevance of two-body 
charge operators from the failure of the impulse approximation (IA) in 
predicting the charge form factors of the three- and four-nucleon 
systems~\cite{Car98}.  
The model commonly used~\cite{Sch90,Car98} includes the $\pi$-, $\rho$-, and 
$\omega$-meson exchange charge operators with both isoscalar and isovector 
components, as well as the (isoscalar) $\rho \pi \gamma$ and (isovector) 
$\omega \pi \gamma$ charge transition couplings, in addition to the 
single-nucleon Darwin-Foldy and spin-orbit relativistic corrections.  
It should be emphasized, however, that for $q \!< \! 5$ fm$^{-1}$ the 
contribution due to the $\pi$-exchange charge operator is typically an order 
of magnitude larger than that of any of the remaining two-body mechanisms 
and one-body relativistic corrections.

We have calculated longitudinal $F_L(q)$ and transverse $F_T(q)$ form factors 
of the $^6$Li ground state as well as transitions from this to the first four 
excited states.
The Coulomb (C$J$) multipoles contributing to $F_L(q)$ are obtained from matrix
elements of the charge operator, $\rho ({\bf q})$, while the electric (E$J$)
and/or magnetic (M$J$) multipoles contributing to $F_T(q)$ are obtained from 
matrix elements of the current operator, ${\bf j}({\bf q})$, using standard 
formulas~\cite{Def66}.

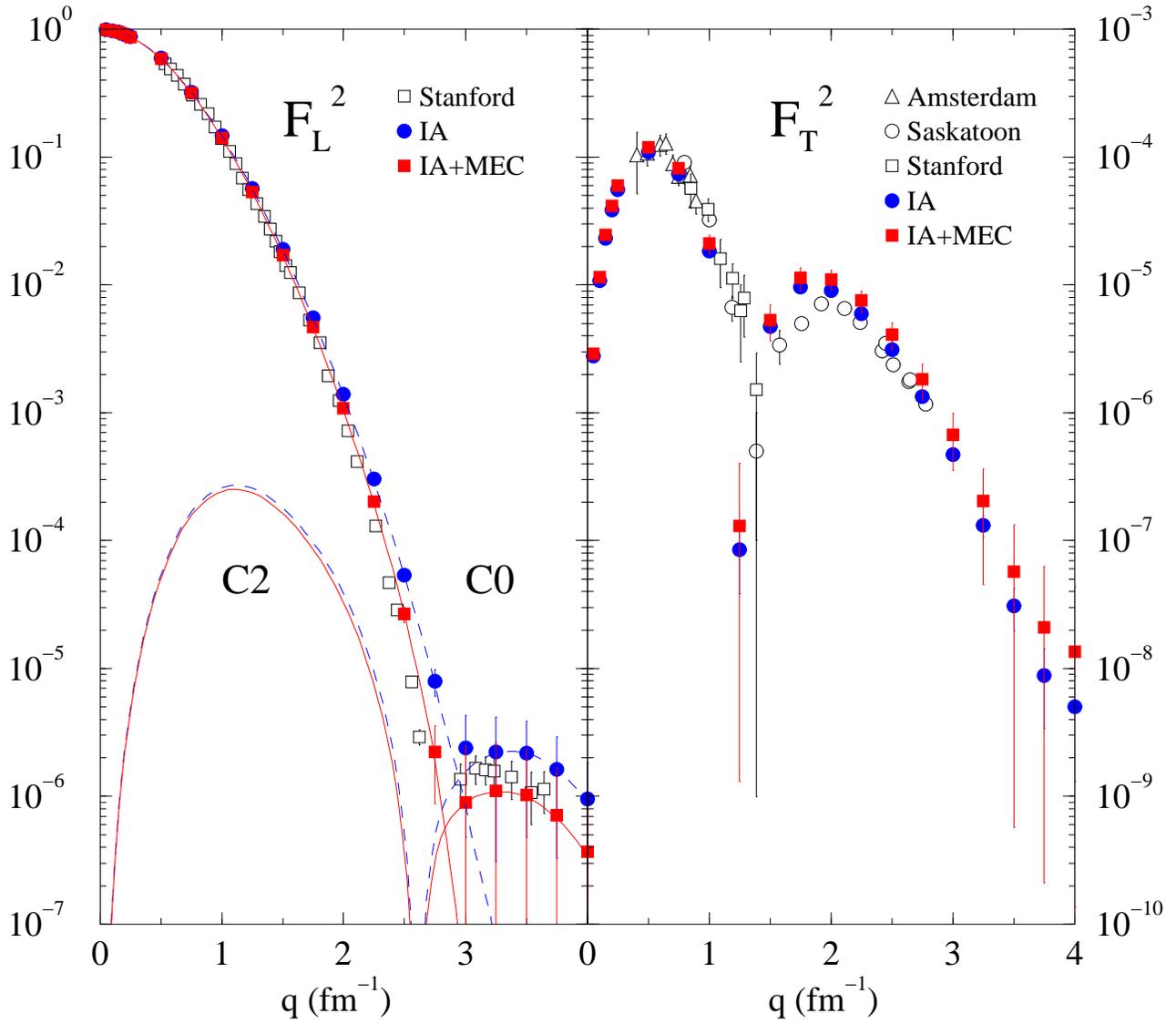
\begin{figure}
\caption{Calculated longitudinal and transverse elastic form factors
of the $^6$Li ground state are shown in impulse approximation (IA) and with
two-body charge and current operators added (IA+MEC) as filled symbols with
Monte Carlo statistical error bars.
The Coulomb monopole (C0) and quadrupole (C2) contributions to the
longitudinal form factor are also shown by the dashed (IA) and solid (IA+MEC)
lines.
Data are from Refs.~\protect\cite{Ber82,Lap78,Li71,Ran66}. }
\label{elastic}
\end{figure}

Our elastic form factors $F_L(q)$ and $F_T(q)$ are compared with
the experimental values~\cite{Ber82,Lap78,Li71,Ran66} in Fig.~\ref{elastic}.  
Since the $^6$Li ground state is (1$^+$;0), both C0 and C2 multipoles
contribute to $F_L(q)$, while only the M1 operator contributes to $F_T(q)$.  
The results obtained in both IA and with inclusion of two-body corrections 
in the charge and current operators (IA+MEC) are displayed, along with the 
statistical errors associated with the Monte Carlo integrations.  
The $F_L(q)$ is in excellent agreement with experiment;
in particular, the two-body contributions (predominantly due to the $\pi$-like 
charge operator) shift the minimum to lower values of $q$, 
consistent with what has been found for the charge form factors
of the hydrogen and helium isotopes~\cite{Car98}.  
The C2 contribution is much smaller than C0 below 3 fm$^{-1}$, as shown in 
Fig.~\ref{elastic}, and at low $q$ is proportional to the ground state 
quadrupole moment.  
Our prediction for the latter is $-0.23(9)$ fm$^2$,
larger (though with a 50\% statistical error) in absolute value than the 
measured value of $-0.08$ fm$^2$, but with the correct (negative) sign. 
Cluster models of the $^6$Li ground state generally give large, positive values 
for the quadrupole moment, presumably due to the lack of D-waves in the 
$\alpha$, and the consequent absence of destructive interference between 
these and the D-wave in the $\alpha$-$d$ relative motion~\cite{Ryz93}. 
For $q \ge 3$ fm$^{-1}$, the C2 contribution becomes dominant, and the shoulder 
seen in the data is entirely due to this component, which has frequently
been omitted from cluster models.

The experimental $F_T(q)$ is well reproduced by our calculations
in the first peak at $q = 0.5$ fm$^{-1}$, but the zero comes a little too
early and the second peak at $q = 2$ fm$^{-1}$ is somewhat overpredicted.
Since the $^6$Li ground state has $T$=0, only isoscalar two-body currents 
contribute to $F_T(q)$; the associated contributions are small at low $q$, 
but increase with $q$, becoming significant for $q > 3$ fm$^{-1}$,  
beyond the range of present data.
The calculated magnetic moment is 0.829 $\mu_N$ in IA and 0.832 $\mu_N$
with two-body currents, about 1\% larger than the experimental value, 
which is close to that of a free deuteron.

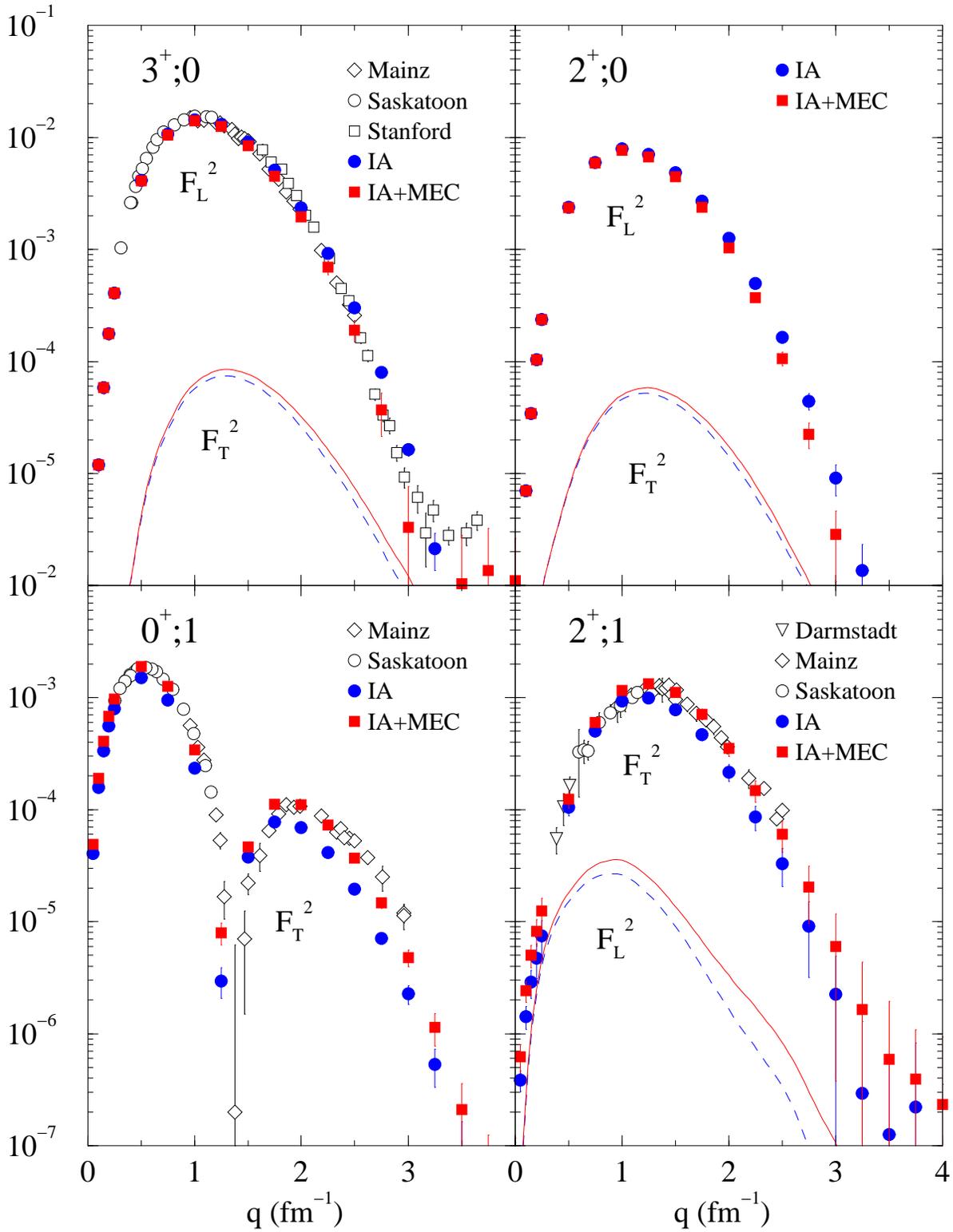
\begin{figure}
\caption{Calculated longitudinal and transverse transition form factors 
to the first four ($J^{\pi};T$) excited states of $^6$Li are shown in impulse
approximation (IA) and with two-body charge and current operators added 
(IA+MEC).
The largest form factor in each case is shown as a point with its statistical 
error bars; the smaller form factor (if any) is shown by dashed (IA) and solid 
(IA+MEC) lines.
Data are from Refs.~\protect\cite{Ber75,Ber79,Li71,Ber76,Deu79,Eig69}. }
\label{transition}
\end{figure}

The calculated longitudinal inelastic form factor to the (3$^+$;0) state is 
found to be in excellent agreement with experiment~\cite{Li71,Ber76,Deu79}, 
as shown in Fig.~\ref{transition}.  
This transition is induced by C2 and C4 operators, 
and thus the associated form factor $F_L(q)$ behaves as $q^4$ at low $q$.  
The two-body contributions only become important for $q > 2$ fm$^{-1}$, but
do improve the agreement with data.
We also show our prediction for the much smaller transverse form factor 
$F_T(q)$.
The calculated radiative width of the (3$^+$;0) state is 
3.38(9)$\times 10^{-4}$ eV in both IA and with MEC, compared to the 
experimental value of (4.40 $\pm$ 0.34)$\times 10^{-4}$ eV~\cite{Eig69}.

Good agreement is also found with the experimental values~\cite{Ber75,Deu79}
for the transverse inelastic form factor to the state (0$^+$;1), 
as shown in Fig.~\ref{transition}.  
This is an isovector magnetic dipole transition and, as expected, is 
significantly influenced, even at low values of $q$, by two-body contributions, 
predominantly by those due to the $\pi$-like current operator.  
The predicted radiative width is 7.49(2) eV in IA and 9.06(7) eV including 
MEC, compared with the experimental value $8.19 \pm 0.17$ eV.
Thus the isovector two-body current contributions increase the $\gamma$-width 
by 20\%.  

The calculated longitudinal and transverse inelastic form factors to the
(2$^+$;0) state are also shown in Fig.~\ref{transition}.
The contributing multipole operators are C2 for the longitudinal
transition, and M1, E2, and M3 for the transverse transition.
The $F_L(q)$ is comparable in magnitude to that for the (3$^+$;0) state,
but has not been measured to date; the $F_T(q)$ is again much smaller.
The corresponding $\gamma$-width is calculated at 8.0(5)$\times 10^{-3}$ eV
both in IA and with MEC.
This result just overlaps the experimental value of (5.4 $\pm$ 2.8)$\times
10^{-3}$ eV~\cite{Eig69}.

Finally, we show the inelastic form factors to the (2$^+$;1) state.
For this isovector transition, the $F_T(q)$ is much larger than the
$F_L(q)$.
The experimental data~\cite{Ber79,Deu79,Eig69} are well reproduced by the 
calculation, with the two-body currents contributing significantly at 
all $q$ values.
However the calculated $\gamma$-width of 0.050(9) eV in IA and 0.075(26) eV
with MEC is several times smaller than the reported experimental
value of 0.27 $\pm$ 0.05 eV.

To summarize, we have presented the first {\it ab initio} microscopic
calculations of $^6$Li elastic and transition form factors, based on six-body
VMC wave functions obtained from realistic interactions and a consistent, 
realistic nuclear electromagnetic current operator.
We do not expect that use of the more accurate GFMC wave functions will lead
to significantly different predictions, since the VMC and GFMC one- and
two-body densities have been found to be quite close~\cite{Pud97}.
Inclusion of the contributions from two-body charge and current operators
brings theory into significantly better agreement with the experimental data.
Thus nuclear many-body theory appears to provide a quantitatively satisfactory
description of the electromagnetic structure of both s- and p-shell
nuclei for a wide range of momentum transfers.
\acknowledgments
 
The authors wish to thank J.\ Carlson, D.\ Kurath, V.R.\ Pandharipande, 
and S.C.\ Pieper for many useful comments.
Computations were performed on the IBM SP of the Mathematics and Computer
Science Division, Argonne National Laboratory.
The work of RBW is supported by the U. S. Department of Energy, Nuclear
Physics Division, under contract No. W-31-109-ENG-38, and that of RS 
by the U. S. Department of Energy.

\begin{table}
\caption{Binding energy, $B$, and excitation energy, $\Delta E$, of $^6$Li 
states in MeV.}
\begin{tabular}{ldddddd}
&\multicolumn{2}{c}{VMC}&\multicolumn{2}{c}{GFMC}&\multicolumn{2}{c}{Expt} \\
$J^{\pi};T$ & $B$ & $\Delta E$ & $B$ & $\Delta E$ & $B$ & $\Delta E$ \\
\tableline
$2^+;1$     & 21.5(1) & 5.5(1) & 25.5(1) & 5.7(1) & 26.62 & 5.37 \\
$2^+;0$     & 22.6(1) & 4.4(1) & 26.8(3) & 4.4(4) & 27.68 & 4.31 \\
$0^+;1$     & 23.2(1) & 3.8(1) & 27.3(1) & 3.9(1) & 28.43 & 3.56 \\
$3^+;0$     & 24.0(1) & 3.0(1) & 28.5(3) & 2.7(3) & 29.80 & 2.19 \\
$1^+;0$     & 27.0(1) & ---    & 31.2(1) & ---    & 31.99 & ---  \\
\end{tabular}
\end{table}

\end{document}